\documentclass{article}

\usepackage[]{graphicx}

\newcommand{\be}{\begin{equation}}
\newcommand{\ee}{\end{equation}}

\begin{document}

% Redefine "plain" pagestyle
%\makeatletter	   % `@' is now a normal "letter' for LaTeX
%\renewcommand{\ps@plain}{%
%     \renewcommand{\@oddhead}{\textrm{Ari Brynjolfsson}\hfil\textrm{\thepage}}% 
%     \renewcommand{\@evenhead}{\@oddhead}%
%     \renewcommand{\@oddfoot}{}% empty footer
%     \renewcommand{\@evenfoot}{\@oddfoot}}
%\makeatother }    % `@' is restored as a "non-letter" character

\title{The type Ia supernovae and the Hubble's constant}         % Enter your title between curly braces
\author{Ari Brynjolfsson \footnote{Corresponding author: aribrynjolfsson@comcast.net}}

        % Enter your name between curly braces
\date{\centering{Applied Radiation Industries, 7 Bridle Path, Wayland, MA 01778, USA}}          % Enter your date or \today between curly braces

\maketitle
%Abstract

\begin{abstract}  The Hubble's constant is usually surmised to be a constant; but the experiments show a large spread and conflicting estimates.  According to the plasma-redshift theory, the Hubble's constant varies with the plasma densities along the line of sight.  It varies then slightly with the direction and the distance to a supernova and a galaxy.  The relation between the magnitudes of type Ia supernovae and their observed redshifts results in an Hubble's constant with an average value in intergalactic space of 59.44 km per s per Mpc.  The standard deviation from this average value is only 0.6 km per s per Mpc, but the standard deviation in a single measurement is about 8.2 km per s per Mpc.  These deviations do not include possible absolute calibration errors.  The experiments show that the Hubble's constant varies with the intrinsic redshifts of the Milky Way galaxy and the host galaxies for type Ia supernovae, and that it varies with the galactic latitude.  These findings support the plasma-redshift theory and contradict the contemporary big-bang theory.  Together with the previously reported absence of time dilation in type Ia supernovae measurements, these findings have profound consequences for the standard cosmological theory.

\end{abstract}



%Table of Contents

% Redefine "plain" pagestyle
\makeatletter	   % `@' is now a normal "letter' for LaTeX
\renewcommand{\ps@plain}{
     \renewcommand{\@oddhead}{\textit{Ari Brynjolfsson: The type Ia supernovae and the Hubble's constant}\hfil\textrm{\thepage}}% 
     \renewcommand{\@evenhead}{\@oddhead}
     \renewcommand{\@oddfoot}{}% empty footer
     \renewcommand{\@evenfoot}{\@oddfoot}}
\makeatother     % `@' is restored as a "non-letter" character

% Set to use the "plain" pagestyle
\pagestyle{plain}

%Section 1: Introduction

\section{Introduction}

We assume that the universe is homogeneous and isotropic on a scale of about $R = c/{\rm{H_0}} \approx 5,000~{\rm{Mpc}} \approx 1.5\cdot 10^{26}~{\rm{m}}.~$  Einstein assumed this "cosmological principle" in his first static model of the universe.  On smaller scales the universe is inhomogeneous and non-isotropic.  Einstein's static model was replaced by a big-bang theory with an explosive event leading to an expansion, which was supposed to supplant Einstein's $\Lambda$ term.  Later, it has been necessary to reintroduce the $\Lambda$ term in form of a time variable dark energy, and to assume dark matter for explaining the observations [1].  In this article, we usually refer to this as the contemporary big-bang theory.  We have shown that plasma redshift, which is based entirely on well proven conventional physics, gives a much simpler explanation of the observations [2,\, 3].  Plasma redshift is an overlooked cross section for energy loss of photons as they penetrate hot sparse plasma.  When considering the plasma-redshift cross-section, there is no need for big bang, dark matter, or time variable dark energy.  Plasma redshift leads to natural explanation of the Hubble's constant and its observed variations, as we show in the following.  Plasma redshift also gives a natural explanation of the highly isotropic cosmic microwave background [2] (see Section 5.9 of that source).  It results in a quasi-static and constantly renewing universe without black holes [2] (see Section 6 of that source).

\indent  The estimates of the Hubble's constant have standard deviations that are much smaller than the spread, usually between $40 ~ {\rm{and}}~ 100 ~ {\rm{km\,s}}^{-1} {\rm{Mpc}}^{-1} ,$ as reported by different observers.  The measurements of the Hubble's constant are often done using objects (stars, Cepheid variables, galaxies, and supernovae) that are at intermediate distances.  The basic distance scales are rooted in parallaxes to nearby stars and Cepheid variables.  These distance scales are then used as measuring rods to galaxies and supernovae.  In the big-bang theory we assume that the distance $R$ is proportional to the redshift, $z,$ and convert the distance scale to a redshift scale.

\indent  According to the plasma-redshift theory, the conversion to redshift scale can lead to errors, because the redshifts per unit distance to the Cepheids in nearby galaxies are relatively large.  The distortion is caused by the higher plasma densities in the coronas of the Milky Way and the host galaxy, and in the space between the galaxies in clusters.  The average electron densities inside the local clusters have higher electron densities between the galaxies than that in the space between the clusters.  The higher electron densities along the lines to nearby galaxies will result in high estimates of the Hubble's constant, $H_0.$  For example, it may be estimated for the nearby galaxies to be about $80 ~ {\rm{km \, s}}^{-1} {\rm{Mpc}}^{-1} ,$ when in the space between the clusters it may be less than $60 ~ {\rm{km \, s}}^{-1} {\rm{Mpc}}^{-1} .$

\indent  We give in Section 2 the basic equations that determine the Hubble's constant and the magnitude-redshift relation.  In Section 3, we compare the equations and the data reported by Riess et al.\,[1].  In Section 4, we analyze the variations in  $H_0.~$ In Section 5, we give the conclusions.

% Section 2: Relations between distances, electron densities, Hubble's constant, redshifts and magnitudes

\section{Relations between distances, electron densities, Hubble's constant, redshifts, and magnitudes}

Brynjolfsson has shown [2] (see in particular Eq.\,(47) of that source) that the plasma redshift cross section for interaction of photons with an electron plasma with electron density $N_e ~ {\rm{cm}}^{-3}$ leads to

\be
{\rm{ln}}\,(1+z) \approx 3.3262 \cdot 10^{-25}\int_{0}^{R} \!\! N_e \, dx ,
\ee  

\noindent  where $z$ is the observed plasma redshift, and $x$ and $R$ are in units of cm.  When the electron density, $N_e ,$ from 0 to $R$ is replaced with its average value, $(N_e)_{\rm{av}},$ along the stretch, we get that

\be
{\rm{ln}}\,(1+z) = 3.3262 \cdot 10^{-25} (N_{e})_{\rm{av}} R ~ \quad {\rm{or}} ~ \quad R = \frac{{3.0064 \cdot 10^{24}}}{{(N_{e})_{\rm{av}}}} \, {\rm{ln}}(1 + z).
\ee

\noindent  For $ R = 1 ~ {\rm{Mpc}} =  3.0851 \cdot 10^{24} ~ {\rm{cm}},$  the redshift is $z = H_0 /c ,$ where the Hubble's constant, $H_0 ,$ is in ${\rm{km \, s}}^{-1} {\rm{Mpc}}^{-1},$ and the velocity of light, $c,$ is in ${\rm{km \, s}}^{-1}.~$  We get then from Eq.\,(2) that 
\[
{\rm{ln}}\,(1+z) \approx z = \frac {{H_{0} }}{{c}} = 3.3262 \cdot 10^{-25} \cdot 3.0851 \cdot 10^{24} \, (N_{e})_{\rm{av}} ,
\] 
or
\be
H_0 = 3.0764\cdot 10^{5}\,(N_{e})_{\rm{av}} \quad {\rm{or}} \quad (N_{e})_{\rm{av}} = 3.2506\cdot 10^{-6}\cdot H_0
\ee
where the Hubble's constant, $H_0 ,$ is in ${\rm{km \, s}}^{-1} {\rm{Mpc}}^{-1},$ and the electron density, $(N_{e})_{\rm{av}},$ is in ${\rm{cm}}^{-3} .$  

\indent  From Eqs.\,(2) and (3), we get when $R$ is in pc, $c$ in ${\rm{km \, s}}^{-1},$ and $H_0 ,$ is in ${\rm{km \, s}}^{-1} {\rm{Mpc}}^{-1},$  that 
\be
R =  \frac {{c \cdot 10^6}}{{H_{0}}} \, {\rm{ln}}\,(1+z)  =  \frac {{2.9979 \cdot 10^{11}}}{{H_{0}}} \, {\rm{ln}}\,(1+z) .
\ee
For small $z$-values, we get the Hubble's law:
\be
R \approx \frac {{2.9979 \cdot 10^{11}}}{{H_{0}}} \, z \, , 
\ee

\indent  The reduction in the light intensity from a star, a supernova, or a galaxy with $R$ is given by
\be
I(R) = \frac{{I_0 \cdot {\rm{exp}}\,\left[\,\, -a_{\rm{av}}\,R - b \,(N_{e})_{\rm{av}} \, R - 2b (N_{e})_{\rm{av}}\, R \right]\,\, }}{{R^2}},
\ee

\noindent  where the factor ${\rm{exp}}\,(-a_{\rm{av}}\,R) $ accounts for the absorption by atoms, molecules and grains along the line of sight, including the absorption in the corona of the Milky Way and the host galaxy.  The factor ${\rm{exp}}\,(- b \,(N_{e})_{\rm{av}})$ accounts for the reduction in light intensity caused by the redshift, and ${\rm{exp}}\,(- 2b \,(N_{e})_{\rm{av}})$ accounts for the Compton scattering in the narrow beam geometry of intergalactic space.  The factor $b = 3.3262 \cdot 10^{-25}$ is the cross section for the plasma redshift, and $2b$ is the cross section for the Compton scattering.

\indent Eq.\,(6) can be simplified with help of the first relation in Eq.\,(2).  We get
\be
I(R) = \frac{{I_0 \cdot {\rm{exp}}\,\left[\,\, -a_{\rm{av}}\,R  \right] }}{{(1+z)^3 \, R^2 }}
\ee
The observed magnitude, m, is defined such that $I(R) = 10^{-0.4\,{\rm{m}}}.~$  The absolute magnitude, M, is defined as the intensity at a distance of 10 pc without any absorbing materials between the observer and the object; that is  $I_0/10^2 = 10^{-0.4\,{\rm{M}}}.~$ When we insert these values into Eq.\,(7) and take the logarithm of the expression, we get
\[
-0.4\, {\rm{m}} \, = \, - 0.4 \, {\rm{M}} + 2 + {\rm{log}} \left({{\rm{exp}}\{\, -a_{\rm{av}}\,R  \}}\right) - 3 \cdot {\rm{log}}\,(1 + z) - 2\cdot{\rm{log}} \, R
\]
or
\be
{\rm{m}} - {\rm{M}} = 5 \cdot{\rm{log}} \, R +7.5 \cdot {\rm{log}}\,(1 + z) - 5 + A \, ,
\ee
where the magnitude, A, of absorption is defined as $ A = - 2.5\cdot {\rm{log}} \left({{\rm{exp}}\{\, -a_{\rm{av}}\,R  \}}\right)\, .~$

\indent  When we for $R$ insert Eq.\,(4) into Eq.\,(8), we get
\be
{\rm{m}} - {\rm{M}} = 5 \cdot{\rm{log}}\, \{{\rm{ln}}\,(1+z)\} +7.5 \cdot {\rm{log}}\,(1 + z) -5 + A - 5 \cdot {\rm{log}}\, (H_{0}) + 5 \cdot {\rm{log}}\, (2.9979 \cdot 10^{11}) \, .
\ee
\noindent  For small redshifts, we can replace the natural logarithm of $(1+z)$ with $z ,$ and get then the Hubble's law.  It is of interest to compare this equation with a similar equation in the big-bang theory, and disregard the deceleration and the acceleration (expansion) terms, see Eq.\,(23) by Sandage [4]: 
\be
{\rm{m}} - {\rm{M}} = 5 \cdot{\rm{log}}\, (z) + 5 \cdot {\rm{log}}\,(1 + z) -5 + A - 5 \cdot {\rm{log}}\, (H_{0}) + 5 \cdot {\rm{log}}\, (2.9979 \cdot 10^{11}) \, .
\ee
\noindent  The second term on the right side consists of two parts, $2.5 \cdot {\rm{log}}\,(1 + z) + 2.5 \cdot {\rm{log}}\,(1 + z) .~$  The first half accounts for the energy loss caused by redshift, and the second half for the intensity loss caused by time dilation (or slower rate of clocks at high velocities).

\indent  The two equations (9) and (10), although fundamentally very different, are numerically almost equal.  When we subtract Eq.\,(9) from (10), we get 
\be
\Delta ({\rm{m}} - {\rm{M}})= - 5 \cdot{\rm{log}}\, \left( \frac{{ {\rm{ln}}\,(1+z) }}{{z}} \right) - 2.5 \cdot {\rm{log}}\,(1 + z) \, .
\ee
The numerical values of Eq.\,(11) are listed in Table 1.  It is seen that the values of the two is nearly the same.  Even at $z=2$, the magnitude value of $\Delta ({\rm{m}} - {\rm{M}})= 0.1081$ is hardly detectable.

%Table1

\begin{table}[h]
\centering
{\bf{Table 1.}} \, \, The variations in $\Delta ({\rm{m}} - {\rm{M}})$ with $z$ as defined by Eq.\,(11).

\vspace{2mm}

\begin{tabular}{llllllll}
\hline
$~z$ & $\Delta ({\rm{m}} - {\rm{M}})$ & $~z$ & $\Delta ({\rm{m}} - {\rm{M}})$ & $~z$ & $\Delta ({\rm{m}} - {\rm{M}})$ & $~z$ & $\Delta ({\rm{m}} - {\rm{M}})$ \\
\hline \hline
0.1 & ~~0.0008 & 0.6 & ~~0.0200 & 1.1 & ~~0.0496 & 1.6 & ~~0.0820 \\
0.2 & ~~0.0030 & 0.7 & ~~0.0254 & 1.2 & ~~0.0560 & 1.7 & ~~0.0885 \\
0.3 & ~~0.0062 & 0.8 & ~~0.0312 & 1.3 & ~~0.0624 & 1.8 & ~~0.0951 \\
0.4 & ~~0.0102 & 0.9 & ~~0.0371 & 1.4 & ~~0.0689 & 1.9 & ~~0.1016 \\
0.5 & ~~0.0149 & 1.0 & ~~0.0433 & 1.5 & ~~0.0754 & 2.0 & ~~0.1081 \\
\hline
\end{tabular}
%\label{}
\end{table}

\indent  The magnitude-redshift relation can usually not be used to see which theory is right.  In the contemporary big-bang theory, it is not reasonable to omit the deceleration and acceleration terms.  These terms can be made to fit almost any magnitude-redshift observations.  In the magnitude-redshift relation defined by the plasma redshift, Eq.\,(9), there is no such flexibility.

% Section 3:  Redshift-magnitude relations and Malmquist bias

\section{Redshift-magnitude relations and Malmquist bias}

Let us assume that the type Ia supernovae (SNe Ia) are perfect standard candles with a well-defined maximum luminosity.  Given these assumptions, would the researchers measuring the magnitudes and the redshifts be able to see which of Eq.\,(9) or (10) is right?  The answer is no, because both equations give answers that within the accuracy of the measurements fit the expectations.  Could we omit the time dilation?  Those who assume that the plasma redshift is correct would say: "There is no time dilation."  The supernova researchers, who are used to the contemporary big-bang theory would say: "No, we cannot omit time dilation, because if we omit the time dilation the SNe Ia would appear brighter than that observed.  The supernova could not be a standard candle, which other indicator show it is."  The supernova researchers, who are unaware of plasma-redshift cross-section, find that because the observations fit the magnitude-redshift relation in the contemporary big-bang theory, the time dilation is required.  It is thus seen that if the supernovae are perfect standard candles, the observations of the magnitude-redshift relation cannot tell us which theory is right.    

\indent  We might, however, use other methods to see which theory is more reasonable [3].  One of them uses the fact that the SNe Ia are imperfect standard candles [5].  The SNe Ia have significant spread in their maximum brightness.  For 111 nearby SNe Ia, Richardson et al.\,[6] found the spread to have a maximum at about ${\rm{M}} \approx -19.46 ,$ with a standard deviation of about $\Delta{\rm{M}} \approx 0.56 .~$  At one standard deviation above, at ${\rm{M}} \approx -19.46-0.56 = -20.02 ,$ the number of nearby SNe Ia falls to about 60.7\,\% of that at the maximum.  Assuming the big-bang theory, Peebles [7] has shown (see in particular Eqs.\, (5.31) and (5.133) of that source), that the number of SNe Ia per redshift interval increases sharply with the redshift, $z .~$  (When using the plasma reshift, the increase would be slightly smaller because of Eq.\,(4) instead of Eq.\,(5)).  When using these equations as a rough guide, we should observe a significant increase in brightness with increasing redshift $z .~$   The fact that Goldhaber et al.\,[8] (see in particular Fig.\,3b of that source) observe no increase whatsoever in the brightness with increasing $z$ indicates that their assumed time dilation, when making the light-curve correction, reduces the absolute-brightness estimates.  This decrease in the estimated brightness is introduced when the supernova researcher adjust the brightness according to the width of the light curve [3].  The same applies to the data reported by Richardson et al.\,[6].

\indent  Brynjolfsson [3] has shown that we get about the right brightness of each SNe Ia, when we correct the absolute magnitude, as determined by the supernova researcher, by the term $\Delta {\rm{M}} = -2.5\, {\rm{log}} \,(1+z).~$  (A slightly better method would be to use the stretch factor, $s',$ obtained by setting the width of the light curve equal to $w = s',$ instead of $s,$ where the measured width is $w = s\,(1+z).\,$)~  We therefore have that the actual absolute magnitude M is
\be
{\rm{M}} = {\rm{M}_{exp}} -2.5\,{\rm{log}}(1 + z),
\ee
\noindent  where ${\rm{M}_{exp}}$ is the absolute magnitude as reported by the supernova researchers, who have assumed that the time dilation applies when they use the light-curve width-correction for estimating the absolute magnitude.  The supernova researchers assumed that the contemporary big-bang theory was correct.  Therefore, they had to use the time dilation for this correction.  The plasma-redshift theory assumes that the universe is quasi-static and has therefore no time dilation, see Eq.\,(9).  The light-curve width-correction, which the supernova researchers applied, resulted in an underestimate of the brightness.  When we correct this underestimate by use of Eq.\,(12), we get an unusually good fit between the observed magnitude-redshift relation and that predicted by the plasma redshift.  For this, see Fig.\,1. 

%Figure1

\begin{figure}[t]
\centering
\includegraphics[scale=.5]{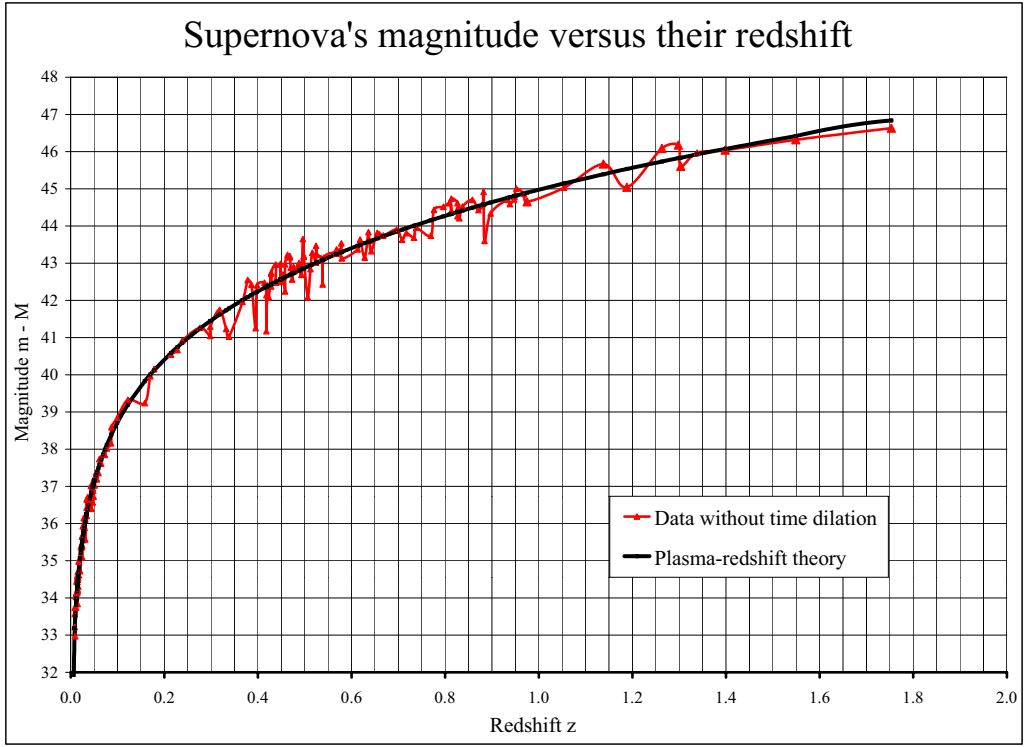}
\caption{The ordinate shows the magnitudes, m-M, of type Ia supernovae versus their redshifts, $z$ from 0.0 to 2, on the abscissa.  The data include all 186 type Ia supernovae reported by Riess et al.\,[1] (see the expanded Tables 5 of that source).  The magnitudes are those derived from Eq.\,(12) with ${\rm{M}_{exp}}$ as reported by Riess et al.\,[1].  The data points indicated with triangles (red) are nearly free of time dilation.  The black curve shows the theoretical predictions of the plasma-redshift theory in accordance with Eq.\,(9), when the a Hubble's constant of ${H}_0 = 59.44 ~{\rm{km\,s}}^{-1}\,{\rm{Mpc}}^{-1}.$}
%\label{}
\vspace{2mm}
\end{figure}

% Section 4:  Spread in the Hubble's constant

\section{Spread in the Hubble's constant}

First, using all the 186 supernovae reported by Riess et al.\,[1], it is required that the average deviation  $\Delta ({\rm{m}}-{\rm{M}}) $ from the theoretical curve, Eq.\,(9), be zero.  The result is $H_0 = 61.6.~$  Second, for the 75 of the nearby, $(z \leq 0.1 ),$ supernovae it is required that $\Delta z$ be determined such that the average deviation of $\Delta ({\rm{m}}-{\rm{M}})~$ from the theoretically expected curve is zero.  The results is $\Delta z = - 0.00185.~$  All $z$-values are then reduced by $ 0.00185.~$  This correction also reduces the standard deviation from $\sigma ({\rm{m}}-{\rm{M}})= 0.306$ to $\sigma ({\rm{m}}-{\rm{M}})= 0.296 .~$  Although this change is insignificant, it shows the trend.  For the 75 SNe Ia, $ z \leq 0.1 ,$ the correction results in 38 above and 37 SNe Ia below the theoretical curve.  The $ 0.00185$-correction corresponds to about $\Delta z = - 0.000925$ for each galaxy.  This reduction in the $z$-values takes into account (roughly eliminates) the average intrinsic redshift, about $\Delta z = 0.000925,$ for each galaxy.  This redshift correction is caused mainly by the coronal plasma in the Milky Way and the host galaxies for the 75 SNe Ia.  This intrinsic redshift correction, although applied to all 186 redshifts, is rather insignificant for high redshift supernovae.  But it lowers the average value of the Hubble's constant for the SNe Ia from 61.60 to 59.44.

\indent  According to Eq.\, (1), the redshift $\Delta z = 0.000925$ corresponds to $\int_{0}^{R} \! N_e \, dx = 2.78\cdot 10^{21} $  ${\rm{cm^{-2}}} .~$  For a thickness of the corona of about 100 kpc, this corresponds to an average electron density of about $(N_e)_{\rm{av}} = 9.0 \cdot 10^{-3}~ {\rm{cm}}^{-3}.~$  Although these corrections are small and nearly insignificant, they are included here because this order of magnitude for the coronal electron density is consistent with many observations [2] (see in particular Section 5.7.1 of that source), which are independent of the present estimates that are derived only from the supernova experiments.  The average electron density in each corona includes the higher densities (because of lower temperatures) in the transition zone to each corona and the densities in the HII-regions within the galaxies.  

\indent   The value of $H_0$ determined in this way for the nearby 75 SNe Ia gives about the same standard deviation, $\sigma ({\rm{m}}-{\rm{M}})= 0.296,$ from the theoretical curve along the entire curve.  It is similar to that about $\sigma ({\rm{m}}-{\rm{M}})= 0.29,$ reported by Riess et al.\,[1] (see Section 5.2 of that source).  In accordance with Eq.\,(3), we find that the standard deviation in $H_0$ is caused partially by a electron density along the line of sight to the supernova.  When the line of sight passes through the corona of another galaxy, or through the higher densities in the space between galaxies in a cluster, the average density is higher.  According to Eq.\,(3), this higher electron density results in larger value of $H_0 .~$  A large Hubble's constant according to Eq.\,(9) results in a smaller value of $({\rm{m}}-{\rm{M}}).~$

\indent  The distribution of the magnitude around the theoretical curve is nearly gaussian for the 186 SNe Ia (the silver as well as gold samples, as listed by Riess et al.\,[1]).  49 samples out of the 186 (or 26.3\,\%) have $|\Delta ({\rm{m}}-{\rm{M}})| \geq 0.3 .~$  If the distribution is gaussian, we should expect about 57.8 samples outside the 0.3 limit.  The standard deviation in 57.8 is about 7.6.  We find therefore that the 49 samples outside the 0.3-limit is consistent with a gaussian distribution.  The supernova researchers select the good samples.  It is likely that other concurrent indicators give reason to discard samples with large deviations.

\indent  Of the 49 samples outside the 0.3-limit, $\mathbf{22}$ (or 45\,\%) have negative deviation with an {\it{average galactic latitude of}} $b = \mathbf{36.2^{\circ}} ,$  and $\mathbf{27}$ (or 55\,\%) have positive deviation with an {\it{average galactic latitude of}} $b = \mathbf{47.5^{\circ}},$ which indicates that the Hubble's constant increases with decreasing latitude, as is to be expected from the plasma redshift theory.  Had we considered only the 111 supernovae that have a redshift  $(z \geq 0.1 ,$ we would have 32 with deviations greater than the 0.3 limit.  The average latitude for the $\mathbf{15}$ with negative deviations is ${b}=\mathbf{37.8^{\circ}} ,$ and the average latitude for the $\mathbf{17}$ with positive deviations is ${b}=\mathbf{49.9^{\circ}} .~$  

\indent  When we limit our investigation to the 75 nearest SNe Ia  $(z \leq 0.1 ),$ 38 had positive and 37 had negative deviation from the theoretically expected value.  Of the 75 SNe Ia, we found 17 (or 23\,\%) with deviation $|\Delta ({\rm{m}}-{\rm{M}})| \geq 0.296 .~$  Of these 17 SNe Ia, 7 (or 41\,\%) have negative deviation, $\Delta ({\rm{m}}-{\rm{M}}) \leq -0.296, $ ) and 10 (59\,\%) have positive deviation.  The average latitude for the $\mathbf{7}$ with negative value is ${b}=\mathbf{32.7^{\circ}} ,$ and the average latitude for the $\mathbf{10}$ with positive value is ${b}=\mathbf{43.5^{\circ}} .~$  
Had we considered only the 45 supernovae that have a redshift  $( 0.1 \geq z \geq 0.0233 ,$ we would have 8 with standard deviations greater than the 0.3 limit.  The average latitude for the $\mathbf{4}$ with negative value is ${b}=\mathbf{36.3^{\circ}} ,$ and the average latitude for the $\mathbf{4}$ with positive value is ${b}=\mathbf{50.7^{\circ}} .~$  

\indent  If we limit our samples to the 30 nearest SNe Ia  nearby, $(z \leq 0.0233 ),$ 19 had positive and 11 had negative deviation.  Of the 30 SNe Ia, we found 9 (or 30\,\%) with deviation $|\Delta ({\rm{m}}-{\rm{M}})| \geq 0.296 .~$  Of these 9 SNe Ia, $\mathbf{3}$ (33\,\%) have negative value with ${b}=\mathbf{27.9^{\circ}}$ and $\mathbf{6}$ (67\,\%) have positive deviation with ${b}=\mathbf{38.7^{\circ}} .~$  

\indent  It is remarkable to find how consistent the latitude effect is independent of the redshift $z.~$  In each case the latitude difference exceeds about ${\Delta b}=\mathbf{10^{\circ}} .~$  A priori, we expect the electron density integral, $\int_{0}^{R} \! N_e \, dx ,$ to increase slightly with decreasing galactic latitude.  As the galactic latitude decreases the integral increases because of the increased path length, mainly through the transition zone to the galactic corona.  We expect this to be a small effect for the Milky Way, because of many other factors affecting the deviations from the theoretical curve.  It is of interest that this galactic latitude effect is clearly supported by the observations.

\indent  The latitude effect caused by absorptions A in Eqs.\,(9) and (10) has been thoroughly documented.  The supernova researchers have done a thorough work at evaluating all corrections.  We have therefore good reason to expect that Riess et al.\,[1] have made these corrections correctly.  We find therefore that not only the average correction, $\Delta z = - 0.00185 ,$ for the average intrinsic redshifts for the Milky Way and the host galaxy for the supernova, but also the above documented latitude effect are consistent with Eq.\,(3), which in turn affects Eq.\,(9) in the observed way.  These effects, which are consistent with the plasma-redshift theory, are inconsistent with the contemporary big-bang theory.

% Section 5:  Conclusions

\section{Conclusions}

In reference [3], we found that the supernova data [1,\,5,\,6,\,8,\,and\,10] show that there is no time dilation.  This contradicts the contemporary big-bang theory, which has time dilation as a basic premise.  The contemporary big-bang theory in addition to big bang and time dilation surmises dark matter and time variable dark energy for explaining the observations.  In contrast the plasma-redshift, an overlooked cross section for interaction of photons with hot, sparse plasma, is based on conventional physics, and requires no such adjustable parameters for explaining the observations, as illustrated in Fig.\,1.  In this paper, we have shown that the experimental data, are consistent with plasma redshift, and indicate an intrinsic redshift for the Milky Way and the host galaxy for each supernova.  This intrinsic galactic redshift, $\Delta z = 0.000925,$ in the Milky Way corresponds to $\int_{0}^{R} \! N_e \, dx = 2.78\cdot 10^{21} $  ${\rm{cm^{-2}}} .~$   This average column density for the electrons derived from the supernova data is consistent with independent observations of the coronal redshifts analyzed in reference [2] (see in particular Section 5.7.1 of that source).  The observations indicate that this intrinsic redshift depends slightly on the galactic latitude.  We have thus a series of observations, all of which contradict the contemporary big-bang theory, but are consistent with the plasma-redshift theory [2].

\indent  The result from the Hubble Space Telescope Key Project [9] found the Hubble constant from type Ia supernovae experiments to be $H_0 = 71 \pm 3_{\rm{random}} \pm 7_{\rm{systematic}}~{\rm{km\,s}}^{-1} {\rm{Mpc}}^{-1} .~$   The supernova data with $z \leq 0.97 $ reported by Riess et al.\,[10] were used in reference [2] without correcting them for the time dilation (see Fig.\,5 of that source).  This resulted in $H_0 = 65.23.~$  But already then these experimental data indicated a need to back-correct them for the faulty time dilation.  When evaluating the experimental data reported by Riess et al.\,[1], with $z \leq 1.755,$ it became necessary to correct them for the faulty time dilation.  In reference [3], the removal of the time dilation from the experimental determination of the magnitude of the supernovae results in them being brighter than that estimated by Riess et al.\,[1].  This together with the expanded data in [1] brings the value of the Hubble's constant from about 65.23 to about $H_0 \approx 61.6~ {\rm{km\,s}}^{-1}\,{\rm{Mpc}}^{-1} .~$  A smaller correction is caused by the intrinsic redshift in the Milky Way and the host galaxy, which reduces the intergalactic value of the Hubble's constant to $H_0 \approx 59.44~ {\rm{km\,s}}^{-1}\,{\rm{Mpc}}^{-1} .~$  The random uncertainty in the average is small or about $\pm 0.6 .~$  To this we must add the uncertainty in the determination of the absolute or systematic value of $H_0,$ which is based mainly on the uncertainty in the Cepheid data.  The evaluation in light of plasma redshift, has already reduced a significant fraction of those systematic errors in $H_0$ that are caused by intrinsic redshifts of the galaxies and galaxy clusters.  It is likely that in the future, we can make better estimates of the electron density along the line of sight and use such estimates for estimating the Hubble's constant, which then can be used to reduce the deviations and uncertainties in the magnitudes of the supernovae, and Cepheid variables.

%References


\begin{thebibliography}{10}

\bibitem{1}    Riess, A. G., Strolger, L.-G., Tonry, J., Casertano, S., Ferguson, H. C., Mobasher, B., Challis, P., Filippenko, A. V., Jha, S., Li, W., Chornock, R., Kirshner, R., Leibundgut, B., Dickinson, M., Livio, M., Giavalisco, M., Steidel, C. C., Ben\'{i}tez, T., \& Tsvetanov, Z. 2004, ApJ, {\textbf{607}}, 665 
\bibitem{2}    Brynjolfsson, A.  2004, arXiv:astro-ph/0401420
\bibitem{3}    Brynjolfsson, A.  2004, arXiv:astro-ph/0406437
\bibitem{4}    Sandage, A. ApJ. 133 (1961) 355
\bibitem{5}    Phillips, M. M. 1993, ApJ, {\textbf{413}}, L108 
\bibitem{6}    Richardson, D., Branch, D., Casebeer, D., Millard, J., Thomas, R. C., \& Baron, E.  2002 AJ, {\textbf{123}}, 745 
\bibitem{7}    Peebles, P. J. E. {\it{Priciples of Physical Cosmology}}, Princeton University Press, ISBN 0-691-01933-9, 1933
\bibitem{8}    Goldhaber, G., Groom, D. E., Kim, A., Aldering, G., Astier, P., Conley, A., Deustua, S. E., Ellis, R., Fabbro, S., Fruchter, A. S., Goobar, A., Hook, I., Irwin, M., Kim, M., Knop, R. A., Lidman, C., McMahon, R. Nugent, P. E., Pain, R., Panagia, N., Pennypacker, C. R., Perlmutter, S., Ruiz-Lapuente, P. Schaefer, B., Walton, N. A., York, T.  2001 ApJ, {\textbf{558}}, 359 
\bibitem{9}    Freedman, W. L., Madore, B. F., Gibson, B. K., Ferrase, L., Kelson, D. D., Sakai, S., Mould, J.R., Kennicutt, Jr., R. C., Ford, H. C., Graham, J. A., Huchra, J. P., Hughes, S. M. G., Illingworth, G. D., Macri, L. M., Stetson, P. B.  2001 ApJ, {\textbf{553}}, 47
\bibitem{10}    Riess, A. G., Filippenko, A. V., Challis, P., Clocchiatti, A. Diercks, A., Garnavich, P. M., Gilliland, R. L., Hogan, C. J., Jha, S., Kirshner, R. P., Leibundgut, B., Phillips, M. M., Riess, D., Schmidt, B. P., Schommer, R. A., Smith, R. C., Spyromilio, J., Stubbs, C., Suntzeff, N. B., \& Tonry, J.  1998, AJ, {\textbf{116}}, 1009







\end{thebibliography}
\end{document}